\documentclass[twocolumn,showpacs,showkeys,superscriptaddress]{revtex4}
\usepackage{amsmath}
\usepackage{amssymb}
\usepackage{graphicx}

\begin{document}

\title{Kantowski-Sachs universes sourced by a Skyrme fluid}

\author{Luca Parisi}
\email{parisi@sa.infn.it}
\affiliation{Dipartimento di Fisica ``E.R. Caianiello", Universit\`{a} di Salerno,\\ Via Giovanni Paolo II  132, 84081 Fisciano (Sa), Italy}
\author{Ninfa Radicella}
\email{ninfa.radicella@sa.infn.it}
\affiliation{Dipartimento di Fisica ``E.R. Caianiello", Universit\`{a} di Salerno,\\ Via Giovanni Paolo II  132, 84081 Fisciano (Sa), Italy}
\affiliation{INFN, Sezione di Napoli, Gruppo Collegato di Salerno, 80126~Napoli, Italy}
\author{Gaetano~Vilasi}
\email{vilasi@sa.infn.it}
\affiliation{Dipartimento di Fisica ``E.R. Caianiello", Universit\`{a} di Salerno,\\ Via Giovanni Paolo II  132, 84081 Fisciano (Sa), Italy}
 \affiliation{INFN, Sezione di Napoli, Gruppo Collegato di Salerno, 80126~Napoli, Italy}

\begin{abstract}
The Kantowski-Sachs cosmological model sourced by a Skyrme field and a cosmological constant is considered in the framework of General Relativity. Assuming a constant radial profile function $\alpha=\pi/2$ for the hedgehog ansatz, the Skyrme contribution to Einstein equations is shown to be equivalent to an anisotropic fluid. Using dynamical system techniques, a qualitative analysis of the cosmological equations is presented. Physically interesting features of the model such as isotropization,  bounce and recollapse are discussed.
\end{abstract}


\maketitle
%%%%%%%%%%%%%%%%%%%%%%%%%%%%%%%%%%%%%%%%%%%%%%%%%%%%%%%%%%%%%%%%%%%%%%%%%%%%%%%%%%%%%%%%%%%%%%%%%%%%%%%%%%%%%%%%%%%%%%%%%%%%%%%%%%%%%%%%%%%%%%%%%%%%%%%%%%%%%%%%%%%%%%%%%%%%%%%%%%%%%%%%%%%%%%%%%%%%%%%%%%%%%%%
\section{Introduction}

The Kantowski-Sachs metrics \cite{Kantowski:1966te} describe spatially homogeneous anisotropic space-times with a four-dimensional isometry group whose three-dimensional subgroup acts multiply transitively on two-dimensional spherically symmetric surfaces (for a clear introduction to the subject see \cite{ELM} and \cite{Vilasi:2002ax}).

The global structure of these models was described by Collins \cite{collins77}, who was also the first who analyzed the model as a two-dimensional dynamical system for the case of perfect fluid with vanishing cosmological constant.

The dynamic of Kantowski-Sachs models has been investigated in the presence of various types of sources such as matter and radiation \cite{Coley:2002vy}, scalar fields  \cite{Reddy:2009zzc,Xanthopoulos:1992fg} and in Einstein-Yang-Mills theory \cite{Donets:1998ts}. Some interesting aspects of these models such as the isotropization \cite{Weber:1984xh,Byland:1998gx,Adhav:2011zzj} and the bouncing behavior \cite{Solomons:2001ef} have also been studied  in detail.

The study of the Kantowski-Sachs models as dynamical systems with compact state space have been presented in \cite{Uggla90} and \cite{Goliath:1998mw}. The dynamical system has been extended to include also a cosmological constant \cite{Goliath:1998na}. For further details on Kantowski-Sachs models and their description through dynamical system theory see \cite{WE} and \cite{CO}. Kantowski-Sachs models have been also considered in theories beyond General Relativity such as String Cosmology \cite{Barrow:1996gx}, extended theories of gravity \cite{Mimoso:1995ge,Shamir:2010ee,Leon:2010pu,Leon:2013bra,Samanta:2013nja}, bimetric theories of gravity \cite{Reddy:2006ysa} and so on.

Moreover, since the metric inside the horizon of a black hole is isometric to a Kantowski-Sachs cosmology, it has been used to study the singularity resolution in loop quantum gravity \cite{Modesto:2004wm,Boehmer:2008fz,Bohmer:2007wi,Joe:2014tca}.

From the observational point of view, the possibility of distinguish between standard Friedmann-RObertson-Walker models and spatially homogeneous but anisotropic models using cosmological data was also considered (see \cite{Mimoso:1993ym,Henriques:1996,Koivisto:2010dr} and references therein). In particular, Kantowski-Sachs cosmologies have been studied in a series of recent papers \cite{Bradley:2011rt,Bradley:2013raa,Keresztes:2013tua} motivated by the observed distribution of inhomogeneities and anisotropies in the cosmic background radiation, and by the possibly different evolution and propagation of perturbations in bouncing and nonbouncing cosmologies.

Recently, self-gravitating Skyrme fields in Kantowski-Sachs gravitational field were considered \cite{Canfora:2013osa}. The Skyrme model is a nonlinear theory of pions. Although not involving quarks, it can be regarded as an approximate, low energy effective theory of QCD. The main motivation for constructing and studying this model is that is has topological soliton solutions that can be interpreted as baryons (Skyrmions). Thus, besides leading to the discovery of new exact analytic solutions of the four-dimensional Skyrme model \cite{Canfora:2013xja,Canfora:2014aia}, these studies have shed light on the bound on the cosmological constant and the bounds on the Skyrme couplings \cite{Canfora:2014jka}, suggesting a possible intrinsic relation between the coupling of the Skyrme field and gravity. For this reason, the dynamics of cosmological models sourced by Skyrme fields is worth being investigated.

In this paper we consider the Einstein-Skyrme system with a cosmological constant in four dimensions as presented in \cite{Canfora:2013osa} and \cite{Canfora:2014jka}. We focus on the Kantowski-Sachs spacetime and assume a constant radial profile function $\alpha=\pi/2$ for the hedgehog ansatz in order to simplify our analysis. This allows us to reduce the field equations to a simple dynamical system which, in spite of its simplicity, shows a physically relevant and interesting behavior.  In particular, two stable fixed points are found, determining the asymptotic  behavior of the whole solution space.

The paper is structured as follows. In Sec. \ref{Cosmeq} the cosmological equations are considered. In Sec. \ref{Dynamical system analysis} the equations are recast as three-dimensional autonomous dynamical system. Relevant dynamical features, namely the stability properties of fixed points, invariant submanifolds and equilibrium sets are analysed. In Sec. \ref{EBI} physically relevant aspects, such as the behavior of exact solutions at fixed points, the conditions for bouncing behavior and isotropization, are discussed. A relation between the cosmological constant and the parameters of the Skyrme model is also found. In Sec. \ref{Conclusions} some conclusions are eventually drawn.

%%%%%%%%%%%%%%%%%%%%%%%%%%%%%%%%%%%%%%%%%%%%%%%%%%%%%%%%%%%%%%%%%%%%%%%%%%%%%%%%%%%%%%%%%%%%%%%%%%%%%%%%%%%%%%%%%%%%%%%%%%%%%%%%%%%%%%%%%%%%%%%%%%%%%%%%%%%%%%%%%%%%%%%%%%%%%%%%%%%%%%%%%%%%%%%%%%%%%%%%%%%%%%%
\section{Cosmological equations} \label{Cosmeq}

The Einstein-Skyrme system with a cosmological constant in four dimensions is described by the total action
\begin{equation*}
S=\frac{1}{16\pi G}\int d^4 x\ \sqrt{-g}\left(R-2\Lambda\right)+S_{Sk},
\end{equation*}
where $R$ is the Ricci scalar, $\Lambda$ represents the cosmological constant, $G$ is the Newton constant and $S_{Sk}$ is the contribution from the Skyrme field. In order to introduce the Skyrme action $S_{Sk}$, let us consider some notations \cite{Manton:2004tk, Nair:2005iw,GVY}.

The Skyrme model is a generalized nonlinear sigma model where the Skyrme field $U$ takes values on a specific target manifold, the Lie group $SU(2)$. The Lie algebra associated to $SU(2)$ will be denoted as $\mathfrak{su(2)}$. Let $\mathfrak{g}$ be a Lie algebra and let $\mathfrak{g}^{*}$ be its dual space (called also the coalgebra). The adjoint representation of $\mathfrak{g}$ is the linear map 
\begin{equation*}
\mathfrak{g} \rightarrow \text{Hom}(\mathfrak{g}, \mathfrak{g}): X \rightarrow \text{ad}_{X}
\end{equation*}
where $\text{ad}_{X} (Y) = [X,Y];\quad X,Y \in \mathfrak{g}$. When $\mathfrak{g}$ is a semisimple Lie algebra, there is a canonical way to identify $\mathfrak{g}$ and $\mathfrak{g}^{*}$. Indeed, the semisimple Lie algebras over $\mathbb{R}$ and $\mathbb{C}$ are characterized by the fact that the Killing form of the algebra $\mathfrak{g}$, is nondegenerate:
\begin{equation*}
 B(X,Y ) = \text{Tr} (\text{ad}_{X} \circ \text{ad}_{Y} );\quad X, Y \in \mathfrak{g}.
\end{equation*}
The trace can be evaluated by taking an arbitrary basis for $\mathfrak{g}$, being the trace independent of the basis choice.

For the considered model, one can define
\begin{eqnarray*}
R^i_{\mu} t_i\equiv R_{\mu}&=&U^{-1}\nabla_\mu U, \\
F_{\mu\nu}&=&\left[R_{\mu},R_{\nu}\right]
\end{eqnarray*}
where the latin indices correspond to the group indices; $t_i=-i\sigma_i$, $\sigma_i$ being the Pauli matrices, i.e. the basis of $\mathfrak{su(2)}$; $R_{\mu}$ being a $\mathfrak{su(2)}$-valued current. The Skyrme action is then defined as
\begin{equation*}
S_{Sk}=\frac{K}{2}\int  d^4 x\ \sqrt{-g} \ \text{Tr}\left(\frac{1}{2}R_{\mu} R^{\mu}+\frac{\lambda}{16} F_{\mu\nu} F^{\mu\nu}\right),
\end{equation*}
where $K$ and $\lambda$ are coupling constants, the latter also involving a dimensionless parameter $e$ introduced by Skyrme to stabilize the solitons \cite{Brown:1995zg}. Both are related to the pion decay constant $F_{\pi}$ as follows: 
\begin{equation*}
K:=F_{\pi}^2/4, \qquad  \lambda:=4/e^2 F^2_{\pi}
\end{equation*}

The Einstein equations acquire a term that can be identified with the energy-momentum tensor $T^{S}_{\mu\nu}$ derived by the variation of the Skyrme action
\begin{equation}
G_{\mu\nu}+\Lambda\ g_{\mu\nu}=8\pi G\  T^{S}_{\mu\nu}, \label{EinEq}
\end{equation}
and the Skyrme equations read
\begin{equation}
\nabla ^{\mu} R_{\mu}+\frac{\lambda}{4} \nabla ^{\mu} \left[R^{\nu},F_{\mu\nu}\right]=0. \label{SkyEq}
\end{equation}
These equations, being nonlinear in nature, are quite difficult to approach. A possible strategy aimed to make the field equations more tractable is to choose a certain ansatz for spherically symmetric systems, the so-called hedgehog ansatz.

Let us first recall the following standard parametrization of the $SU(2)$-valued scalar field $U$:
\begin{equation*}
U(x_\mu) = Y^{0} \mathbb{I}+Y^{i}t_{i},\qquad U^{-1}(x_\mu) = Y^{0} \mathbb{I}-Y^{i}t_{i},
\end{equation*}
where $Y^{0}= Y^{0}(x_\mu)$ and $Y^{i}= Y^{i}(x_\mu)$ satisfy
\begin{equation*}
(Y^{0})^{2} + Y^{i}Y_{i} = 1.
\end{equation*}

The name hedgehog derives from the fact that the fields of this configuration point radially outward from the origin of the inner space at all points in space-time. In terms of the group element $U$, the hedgehog ansatz reads
\begin{equation*}
U= \mathbb{I} \cos \alpha + n^{i} t_{i} \sin \alpha, \qquad U^{-1} = \mathbb{I} \cos \alpha - n^{i} t_{i} \sin \alpha,
\end{equation*}
where $n^{i}$ ($i = 1, 2, 3$) are given by
\begin{equation*}
n^{1} = \sin \theta \cos \phi, \quad n^{2} = \sin \theta \sin \phi, \quad n^{3} = \cos \theta.
\end{equation*}
The function $\alpha$ is the so-called radial profile function. When dealing with spherically symmetric space-times, $\alpha$ depends on the coordinates of a two dimensional Lorentzian manifold, namely $\alpha=\alpha(y)$ with $y^{A}$, $A=0,1$. In terms of the variables $Y^{0}$ and $Y^{i}$, the ansatz corresponds to
\begin{equation*}
Y^{0} = \cos \alpha, \qquad Y^{i} = n^{i} \sin \alpha.
\end{equation*}
All the above-defined quantities can be expressed in terms of the radial profile function and the metric functions. Eventually the hedgehog ansatz allows to reduce the Skyrme equations in Eq.(\ref{SkyEq}), that is a system of coupled nonlinear partial differential equations, to a single scalar equation  (see \cite{Canfora:2013osa} for step-by-step calculations).

In what follows we consider the additional restriction for the metric to be the Kantowski-Sachs one, which reads
\begin{equation*}
ds^2=-dt^2+A(t)^2\ dr^2+B(t)^2\left[d\theta^2+\sin{\theta}^2 d\phi^2\right].
\end{equation*}
Moreover, we consider the particular case of a constant radial profile function $\alpha=\pi/2$ for which the scalar Skyrme equation is identically solved. We remark that, in spite of its simplicity, this solution is not trivial, because it actually affects the gravitational equations of motion through a nonvanishing and nonconstant energy momentum tensor. Under these hypotheses, once the stress-energy tensor is expressed in terms of the metric functions, Eq.(\ref{EinEq}) eventually reads
\begin{eqnarray}
2 \frac{\dot{B}\dot{A}}{BA}+\frac{1}{B^2}+\frac{\dot{B}^2}{B^2}-\Lambda&=&8\pi G \left[\frac{K}{B^2}\left(1+\frac{\lambda}{2B^2}\right)\right]\label{eq.00}\\
2\frac{\ddot{B}}{B}+\frac{1}{B^2}+\frac{\dot{B}^2}{B^2}-\Lambda&=&8\pi G\left[ \frac{K}{B^2}\left(1+\frac{\lambda}{2B^2}\right)\right]\label{eq.11}\\
\frac{\dot{B}\dot{A}}{BA}+\frac{\ddot{B}}{B}+\frac{\ddot{A}}{A}-\Lambda&=&-8\pi G\left[ \frac{K\lambda}{2B^4}\right].\label{eq.22}
\end{eqnarray}
In the equations above the dot represents derivation with respect to time. The Skyrme parameters $F_\pi$ and $e$ are fixed by fitting the energies of a quantized Skyrmion to the masses of the nucleon and $\Delta$ resonance. From flat space-time results \cite{adkins83}, one gets $8\pi G\ K\sim 1.5\cdot 10^{-39}$ and $\lambda\sim  2\cdot 10^{-31} m^2$. In this particular case the parameter $K$ acts as a rescaling of the Newton constant: $G_{eff}= G  K$; therefore, we henceforth set  $8\pi G K \rightarrow k$. Motivated by the numerical evaluation above, in what follows we will consider $0<k<1$.

A close inspection of the equations reveals that the contribution of the Skyrme action to the Einstein equations traces an anisotropic fluid. In particular, the Skyrme field behaves as a fluid with different radial and tangential pressures, whose energy momentum tensor can be written as follows:
\begin{equation}
T_{\mu\nu}=(\rho+p_t)u_\mu u_\nu+p_t g_{\mu\nu}+(p_r-p_t)\chi_\mu\chi_\nu \label{SET},
\end{equation}
where $u_{\mu}$ is the four-velocity and $\chi^\mu$ a unit spacelike vector in the radial direction, i.e. $\chi^\mu=A^{-1} \delta^\mu_r$. Moreover $\rho$ represents the energy density, $p_r$ the radial pressure measured in the direction of $\chi^\mu$ and $p_t$ the transverse pressure measured in the orthogonal direction to $\chi^{\mu}$. Then, by comparison with Eqs.(\ref{eq.00})-(\ref{eq.22}), the quantities appearing in Eq.(\ref{SET}) read
\begin{eqnarray}
\rho&=&\frac{1}{B^2}\left(1+\frac{\lambda}{2B^2}\right)\label{rho}\\
p_r&=&-\rho\label{pr}\\
p_t&=&\omega_t\rho, \quad \omega_t =-1+\frac{2(\lambda+B^2)}{\lambda+2B^2}\label{pt}.
\end{eqnarray}
We remark that this analogy is only valid under the considered hypotheses and it should be possible to distinguish between the Skyrme field and an anisotropic fluid by taking into account linear perturbations. Nevertheless, this analogy allows us to draw a parallel that will be helpful in the following analysis.

Such an anistropic fluid satisfies all the energy conditions - weak, strong and dominant energy condition \cite{lobo06, chan09}. Let us now consider the behavior of the cosmological constant; it violates the strong energy condition, since it gives a positive contribution for the variation of the expansion of the geodesics curves in the congruence, in contrast to the convergency effect of matter. When we consider this additional effect, the total fluid, composed by both the Skyrme fluid and the cosmological constant, could violate or fulfill the strong energy condition depending on the relative contribution of the pressure terms. In order to verify the strong energy condition, the positive tangential pressure of the Skyrme fluid must compensate the negative pressure of the cosmological constant.
\begin{equation}
 p_t \geq \Lambda.
\end{equation} 
 
 %%%%%%%%%%%%%%%%%%%%%%%%%%%%%%%%%%%%%%%%%%%%%%%%%%%%%%%%%%%%%%%%%%%%%%%%%%%%%%%%%%%%%%%%%%%%%%%%%%%%%%%%%%%%%%%%%%%%%%%%%%%%%%%%%%%%%%%%%%%%%%%%%%%%%%%%%%%%%%%%%%%%%%%%%%%%%%%%%%%%%%%%%%%%%%%%%%%%%%%%%%%%%%%
\section{Dynamical system analysis} \label{Dynamical system analysis}

It is convenient to recast the equations as an autonomous system of first order nonlinear differential equations and then to perform a local analysis to characterize the stability of the stationary points corresponding to specific cosmological solutions. In this particular model, as will be shown below, due to the introduction of suitably defined variables, it is possible to perform a compactification of the phase space gaining informations on the behavior of the model at infinity.

\begin{figure}[h!]
\centering
 \includegraphics[scale=0.70]{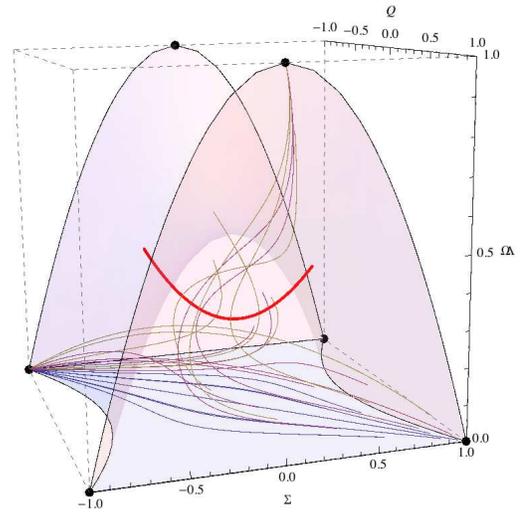}
\caption{{\small Phase space of the system in Eqs. (\ref{sistema3d1})-(\ref{sistema3d3}). We have arbitrarily set $k=0.5$. The orbits corresponding to physical solutions are bounded in the region delimited by the three submanifolds $Q=1$, $Q=-1$, $\Omega_\Lambda=0$} and the surface arising from the constraints. Only six out of the eight fixed points are depicted, the remaining two being outside of the allowed region of physical solutions. The thick (red) curve is an arc of a parabola corresponding to the normally hyperbolic equilibrium set.}
\label{3D}
\end{figure}

The Einstein field equations can be written in terms of propagation equations for the usual volume expansion scalar $\theta$, the shear scalar
$\sigma^2=\frac{1}{2}\sigma^{\mu\nu}\sigma_{\mu\nu}$ (where $\sigma^{\mu\nu}$ is the shear tensor) and the 3-curvature scalar   ${^{\scriptscriptstyle{(\!3\!)}}\!R}$ which, for the Kantowski-Sachs metric, are:
\begin{eqnarray}
{^{\scriptscriptstyle{(\!3\!)}}\!R}=\frac{2}{B^2}\\
\sigma=\frac{1}{ \sqrt{3}} (\frac{\dot{A}}{A} -\frac{\dot{B}}{B})\\
\theta= \frac{\dot{A}}{A} +2\frac{\dot{B}}{B}.
\end{eqnarray}
The Friedmann equations become\begin{eqnarray}
\dot{\theta}+\frac{1}{3}\theta^2+2\sigma^2&=&\Lambda-\frac{k\lambda {^{\scriptscriptstyle{(\!3\!)}}\!R^2}}{8}\\
\dot{\sigma}+\theta\sigma-\frac{1}{2\sqrt{3}}{^{\scriptscriptstyle{(\!3\!)}}\!R}&=&-\frac{k{^{\scriptscriptstyle{(\!3\!)}}\!R}}{4 \sqrt{3}}\left(\lambda {^{\scriptscriptstyle{(\!3\!)}}\!R}+2\right)\\
{^{\scriptscriptstyle{(\!3\!)}}\!\dot{R}}+\frac{2}{3}\theta {^{\scriptscriptstyle{(\!3\!)}}\!R} -\frac{2}{\sqrt{3}}{^{\scriptscriptstyle{(\!3\!)}}\!R}\sigma&=&0\\
{^{\scriptscriptstyle{(\!3\!)}}\!R}+\frac{2}{3}\theta^2-2\sigma^2&=&2\Lambda+k {^{\scriptscriptstyle{(\!3\!)}}\!R}\!\left(\!1+\frac{\lambda {^{\scriptscriptstyle{(\!3\!)}}\!R}}{4}\right)\!\!. \label{vincolo}
\end{eqnarray}
The quantity $\frac{1}{9}\theta^2+\frac{1}{6} {^{\scriptscriptstyle{(\!3\!)}}\!R}$ is strictly positive even if $\theta=0$; this means that the new variable $D\equiv\sqrt{\frac{1}{9}\theta^2+\frac{1}{6} {^{\scriptscriptstyle{(\!3\!)}}\!R} }$ is a well-defined normalization. From Eq.(\ref{vincolo}) one gets the following constraint:
\begin{equation} \label{vincoli}
D^2=\frac{1}{3}\left[\sigma^2+\Lambda+\frac{k{^{\scriptscriptstyle (\!3\!) }\!R}}{2}+\frac{k\lambda {^{\scriptscriptstyle (\!3\!) }\!R^{2}}}{8}\right].
\end{equation}
It is then possible to introduce new dimensionless variables
\begin{eqnarray*}
&Q=\frac{\theta}{3D},\quad\Sigma^2=\frac{\sigma^2}{3D^2},\quad\Omega_\Lambda=\frac{\Lambda}{3D^2},\\
&\Omega_k=\frac{ {^{\scriptscriptstyle{(\!3\!)}}\!R}}{6D^2} \quad\text{and} \quad \Omega_s=\frac{k\lambda {^{\scriptscriptstyle{(\!3\!)}}\!R^2}}{24 D^2}.
\end{eqnarray*}
This allows to construct a compact state space since the constraint in Eq.(\ref{vincoli}) becomes
$$
Q^2+\Omega_k=1,\quad\quad\Omega_\Lambda +\Sigma^2+k(1-Q^2)+\Omega_s=1.
$$
Then one can have a complete picture of the cosmological behavior once one introduces a normalized time derivative $'\equiv \frac{d}{d\tau}=\frac{1}{D}\frac{d}{dt}$. Since $D$ is real valued and strictly positive, it provides a monotonically increasing time variable. Deriving all the variables with respect to $\tau$, the equations of motion become
\begin{eqnarray}
\theta'\!\!\!&=&\!\!\!3D\left[-\Sigma^2+2\Omega_\Lambda-Q^2(k+1)+(k-1)\right]\\
\sigma'\!\!\!&=&\!\!\! \sqrt{3}D \!\left[(1\!-\!Q^2)(1\!+\!k)\!-\!2+2\Omega_\Lambda\!-\! 3Q\Sigma +2\Sigma^2\right]\\
R'\!\!\!&=&\!\!\!12 D^2(1-Q^2)(\Sigma-Q).
\end{eqnarray}
One also gets
\begin{equation}
D'=\frac{1}{3D}\left(\frac{\theta\theta'}{3}+\frac{1}{4}R'\right).
\end{equation}
The dynamical system can then be recast in the following form:
\begin{eqnarray}
Q'&=&Q\left(\frac{\theta'}{\theta}-\frac{D'}{D}\right),\\
\Omega'_\Lambda&=&-2\Omega_\Lambda\frac{D'}{D},\\
\Sigma'&=&\Sigma\left(\frac{\sigma'}{\sigma}-\frac{D'}{D}\right),\\
\Omega'_k&=&2 \Omega_k\left(\Sigma-Q-\frac{D'}{D}\right),\\
\Omega'_s&=&2\Omega_s\left[2(\Sigma-Q)-\frac{D'}{D}\right].
\end{eqnarray}
Eventually, making use of the constraint in Eq.(\ref{vincoli}), the system is reduced to a three-dimensional autonomous dynamical system in the new variables $Q, \Sigma, \Omega_\Lambda$:
\begin{eqnarray}
Q'\!\!\!&=&\!\!\!(Q^2 - 1) (1 - k (1 - Q^2) + Q \Sigma + \Sigma^2 - 2 \Omega_\Lambda)  \label{sistema3d1} \\
\Sigma'\!\!\!&=&\!\!\!k (1 - Q^2)(1 - Q \Sigma )- (1-\Sigma ^2) \left[ 1+ Q(Q+\Sigma) \right] \nonumber \\
&&+2 (1-Q \Sigma) \Omega_\Lambda \label{sistema3d2} \\
\Omega_\Lambda'\!\!\!&=&\!\!\!2 \left[Q\left(2\! - \! k(1\! - \! Q^2)\right)\!\!+\!\!\Sigma(Q \!+\!\Sigma\!-\!1) -2\Omega_\Lambda \right] \Omega_\Lambda  \label{sistema3d3}
\end{eqnarray}
The system has a compact phase space defined as follows:
\begin{eqnarray}
\mathcal{S} &=& \{ (Q, \Sigma, \Omega_{\Lambda}) \in \mathbb{R} \ | \ -1\le Q	\le 1,  -1 \le \Sigma	\le 1,    \nonumber \\
&&  0 \le \Omega_{\Lambda} \le 1, 0 \le 1 - \Omega_{\Lambda} - \Sigma^2 - k(1 - Q^2) \le 1 \} \nonumber
\end{eqnarray}
An example is depicted in Fig.\ref{3D}.

The system in Eqs.~(\ref{sistema3d1})-(\ref{sistema3d3}) admits eight stationary points listed in Table \ref{tab1}.
\begin{table}[h]
\begin{center}
\begin{tabular}{|c|c|c|c|c|c|}
\hline
Sol.&  $Q$   & $\Sigma$ & $\Omega_\Lambda$ & $\Omega_K$ & $\Omega_S$ \\  \hline\hline
 $\mathcal{A}$  & $-1$    & $-1$ & $0$ & $0$ & $0$ \\
 $\mathcal{B}$  & $-1$    & $0$  & $1$ & $0$ & $0$ \\
 $\mathcal{C}$  & $-1$    & $1$  & $0$ & $0$ & $0$ \\
 $\mathcal{D}$  & $-p$  & $-p$ & $1$ & $s$ & $q$\\
 $\mathcal{E}$  & $p$  & $p$ & $1$ & $s$ & $q$ \\
 $\mathcal{F}$  & $1$  & $-1$& $0$ & $0$ & $0$ \\
 $\mathcal{G}$  & $1$  & $0$ & $1$  & $0$ & $0$ \\
 $\mathcal{H}$  & $1$  & $1$ & $0$  & $0$ & $0$ \\ \hline
\end{tabular}
\caption{Stationary points for the system in Eqs.(\ref{sistema3d1})-(\ref{sistema3d3}) with $p=\sqrt{\frac{1 + k}{2 + k}}$, $s=\frac{1}{2+k}$ and $q=-\frac{1 + 2 k}{2 + k}$. For the sake of completeness we have also reported the corresponding values of $\Omega_K$ e $\Omega_S$.}
\label{tab1}
\end{center}
\end{table} 
The eigenvalues of the Jacobian matrix evaluated at the equilibrium points allow to characterize their stability; the results are listed in Table \ref{tab2}. The points $\mathcal{A}$ and $\mathcal{G}$ are attractors. The point $\mathcal{B}$ and the point $\mathcal{H}$ are repellers. The points $\mathcal{C, D, E, F}$ are unstable of the saddle type having at least two eigenvalues with real part of opposite signs. It is worth stressing that the two points  $\mathcal{D}$ and  $\mathcal{E}$  are always placed outside the physical region $\mathcal{S}$ of the phase space.

\begin{table}[h]
\begin{center}
\begin{tabular}{|c|c|c|c|c|}
\hline
Point &  Stability & $\lambda_1$ & $\lambda_2$& $\lambda_3$    \\  \hline\hline
 $\mathcal{A}$  & Stable (attractor)   & $-6$ & $-6$ & $-6$  \\
 $\mathcal{B}$  & Unstable (repeller)  & $4$  & $3$ & $2$  \\
 $\mathcal{C}$  & Unstable (saddle)    & $-6$  & $-2$ & $2$  \\
 $\mathcal{D}$  & Unstable (saddle)    & $0$ & $<0$ & $>0$  \\
 $\mathcal{E}$  & Unstable (saddle)    & $0$  & $<0$ & $>0$ \\
 $\mathcal{F}$  & Unstable (saddle)    & $6$  & $-2$ & $2$  \\
 $\mathcal{G}$  & Stable (attractor)   & $-4$ & $-3$  & $-2$  \\
 $\mathcal{H}$  & Unstable (repeller)  & $6$ & $6$  & $6$  \\ \hline
\end{tabular}
\caption{Stability of the stationary points in Table \ref{tab1}. In the last three columns the sign of the real part of the eigenvalues for the linearized system is represented.}
\label{tab2}
\end{center}
\end{table}

The system also displays a curve of equilibrium points lying in the $Q=\Sigma$ plane and is defined by:
\begin{equation}
-\sqrt{\frac{1 - k}{4 - k}} \leq Q \leq \sqrt{\frac{1 - k}{4 - k}}, \quad \Omega_\Lambda=\frac{1}{2} (1 + 2 Q^2 + k ( Q^2-1)) \label{equiset}
\end{equation}
The Jacobian matrix evaluated at the points of the equilibrium set has two real eigenvalues of opposite signs and a third, vanishing, eigenvalue [except for the point with $Q=\Sigma=0$ and $\Omega_\Lambda=(1-k)/2$, which has vanishing eigenvalues requiring further investigation]. Thus it is a {\it normally hyperbolic} equilibrium set.

\begin{figure}[h!]
\centering
 \includegraphics[scale=0.80]{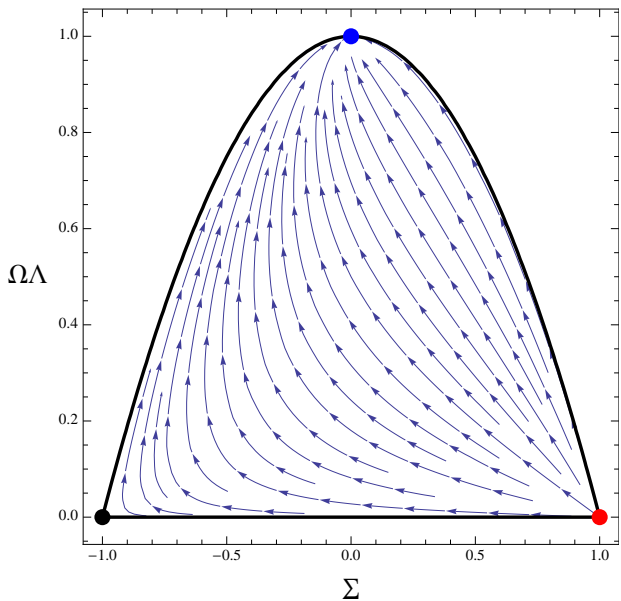}
\caption{{\small In the submanifold  $Q=1$ there are three equilibrium points corresponding to the points $\mathcal{F}$, $\mathcal{G}$ and $\mathcal{H}$ of Table \ref{tab1}. In this submanifold, $\mathcal{F}$  (down-left, black) is unstable of the saddle type;  $\mathcal{G}$ (top-center, blue) is stable and $\mathcal{H}$ (down-right, red) is unstable. Thus, in this subspace, $\mathcal{G}$ is a future attractor.}}
\label{QU1}
\end{figure}

\begin{figure}[h!]
\centering
 \includegraphics[scale=0.80]{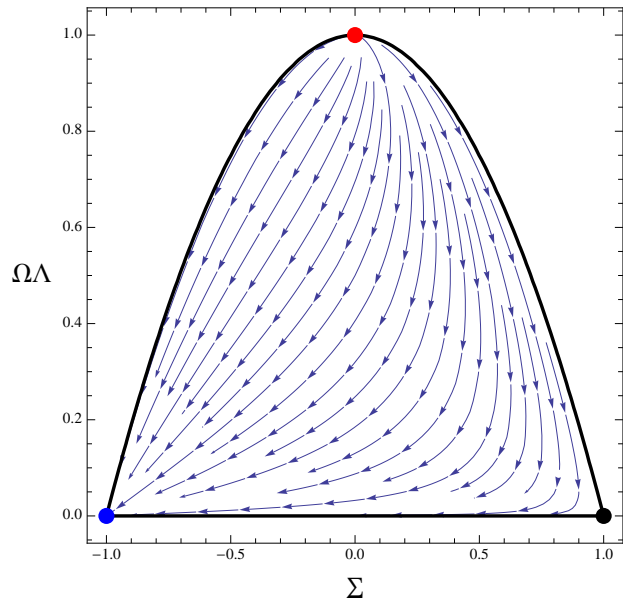}
\caption{{\small In the submanifold $Q=-1$, there are three equilibrium points corresponding to the points $\mathcal{A}$, $\mathcal{B}$ and $\mathcal{C}$ of Table \ref{tab1}. In this submanifold, $\mathcal{A}$  (down-left, blue) is stable;  $\mathcal{B}$ (top-center, red) is unstable and $\mathcal{C}$ (down-right, black) is unstable of the saddle type. Thus, in this subspace, $\mathcal{A}$ is a future attractor.}}
\label{QU2}
\end{figure}

\begin{figure}[h!]
\centering
 \includegraphics[scale=0.80]{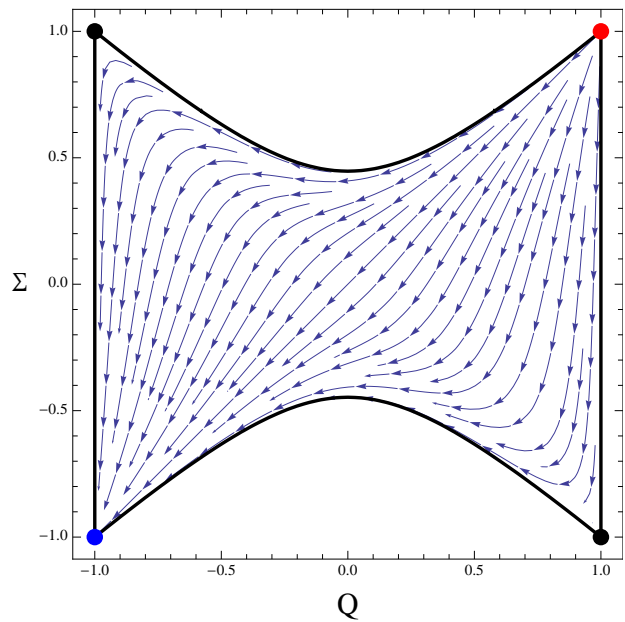}
\caption{{\small In the submanifold $\Omega_\Lambda=0$, there are four equilibrium points corresponding to the points $\mathcal{A}$, $\mathcal{C}$, $\mathcal{F}$ and $\mathcal{H}$ of Table \ref{tab1}. The points $\mathcal{C}$ (top-left, black) and $\mathcal{F}$ (down-right, black) are unstable of the saddle type; $\mathcal{H}$  (up-right, red) is unstable; $\mathcal{A}$ (down-left, blue) is stable. Thus, in this subspace, $\mathcal{A}$ is a future attractor.}}
\label{OLZ}
\end{figure}

The system in Eqs. (\ref{sistema3d1})-(\ref{sistema3d3}) has three invariant submanifolds characterized by $Q=1$, $Q=-1$ and  $\Omega_\Lambda=0$, depicted in Figs.\ref{QU1}, \ref{QU2} and \ref{OLZ}.

%%%%%%%%%%%%%%%%%%%%%%%%%%%%%%%%%%%%%%%%%%%%%%%%%%%%%%%%%%%%%%%%%%%%%%%%%%%%%%%%%%%%%%%%%%%%%%%%%%%%%%%%%%%%%%%%%%%%%%%%%%%%%%%%%%%%%%%%%%%%%%%%%%%%%%%%%%%%%%%%%%%%%%%%%%%%%%%%%%%%%%%%%%%%%%%%%%%%%%%%%%%%%%%
\section{Exact solutions at fixed points, bounces, isotropization} \label{EBI}

It is interesting to test whether the dynamics described by the system Eqs.(\ref{eq.00})-(\ref{eq.22}) leads to physically relevant conditions such as isotropization. We first show the procedure which allows to reconstruct the time evolutions of the two scale factors at the fixed points. Then we consider the conditions allowing bouncing solutions. Eventually, isotropization is discussed. 
 
\subsection{Solution reconstruction} 
A first step toward the physical interpretation of the found solutions is the analysis of the metric functions at the fixed points \cite{Goliath:1998mw,Goheer:2007wu}. The evolution of the two scale factors $A$ and $B$ can be reconstructed as follows. The Raychauduri equation can be rewritten in terms of the dimensionless variables:
\begin{equation}\label{ray2}
\dot{\theta}=-\left(1+\frac{2\Sigma^2}{Q^2}-\frac{\Omega_\Lambda}{Q^2}+\frac{\Omega_s}{Q^2}\right)\frac{\theta^2}{3}
\end{equation}
and then can be evaluated for each fixed point to give the analytic behavior of the volume expansion scalar $\theta$. It is useful to rewrite Eq.(\ref{ray2}) in terms of a deceleration parameter
\begin{equation}
q=\frac{2\Sigma^2}{Q^2}-\frac{\Omega_\Lambda}{Q^2}+\frac{\Omega_s}{Q^2}\quad\Longrightarrow\quad \dot{\theta}=-\left(1+q\right)\frac{\theta^2}{3}.
\end{equation}
Then, for each fixed point one can evaluate the corresponding $q$. Two classes of solution are obtained. For the stationary points $\mathcal{A}$, $\mathcal{C}$, $\mathcal{F}$ and $\mathcal{H}$, one gets 
\begin{equation}
q=2,\quad\theta\sim t^{-1}.
\end{equation}
These points are characterized by $Q^2=1$ and $\Sigma^2=1$; this allows to solve in terms of both scale factors $A$ and $B$ to obtain either
\begin{equation}
B\sim \text{const.}\quad\text{and}\quad A\sim t,
\end{equation}
or
\begin{equation}
B\sim t^{2/3} \quad\text{and}\quad A\sim t^{-1/3} ,
\end{equation}
depending on the sign of $\Sigma$. Thus, these points represent Kasner-like solutions.

For the two points that have zero shear and $\Omega_\Lambda=1$, namely $\mathcal{B}$ and $\mathcal{G}$, one gets
\begin{equation}
q=-1,\quad\theta\sim\text{const}.
\end{equation}
A vanishing shear implies the same evolution for both $A$ and $B$ that is driven by the cosmological constant, the sign of the exponent depending on the sign of $Q$:
\begin{equation}
B\sim A \sim e^{\pm\sqrt{\frac{\Lambda}{3}} t},
\end{equation}
thus these points represent de Sitter-like solutions. Analogously, for the equilibrium set Eq.(\ref{equiset}) the acceleration parameter is always $q=-1$.

One can easily check that these results are consistent with those found in \cite{Goliath:1998na} for the vacuum boundary. Indeed, setting $k=0$ and $\Omega_s=0$, that is, for a vanishing Skyrme source terms, the constraints become
\begin{eqnarray*}
&&\Omega_k=1-Q^2,\\
&&\Omega_\Lambda=1-\Sigma^2.
\end{eqnarray*}
and the system can be expressed in terms of the variables $Q$ and $\Sigma$ (remembering also that, in this correspondence,  $\Sigma \rightarrow -Q_{+}$)). The correspondence is straightforward and reads as shown in Table \ref{tab3}.
\begin{table}[htb]
\begin{tabular}{|c|c|c|c|c|c|c|}
\hline
Skyrme        & $\mathcal{A}$& $\mathcal{B}$& $\mathcal{C}$  & $\mathcal{F}$& $\mathcal{G}$& $\mathcal{H}$ \\
\hline 
Perfect fluid & $_{-}K_{+}$  &     $_{-}dS$ & $_{-}K_{-}$       & $_{+}K_{+}$  & $_{+}dS$     & $_{+}K_{-}$ \\
\hline
\end{tabular}
\caption{Correspondence between fixed point in \ref{tab2} and those  found in \cite{Goliath:1998na}.}\label{tab3}
\end{table}

\subsection{Bounce and recollapse}

The motivation to consider bouncing behavior in Kantowski-Sachs models is twofold \cite{Solomons:2001ef}. First, the geometry of the universe at a bounce might be different from the isotropic and spatially homogeneous Friedmann-Robertson-Walker spacetimes. Second, one might expect the Kantowski-Sachs geometry, having the same symmetries as the spatially homogeneous interior region of the extended (vacuum) Kruskal solution, suitable for describing the turning point in black hole collapse and subsequent expansion.

Following \cite{Solomons:2001ef}, let us define an expansion parameter for each scale factor
\begin{equation}
x=\frac{\dot{B}}{B}\quad\text{and}\quad y=\frac{\dot{A}}{A}.
\end{equation}
A bounce in the scale factor $A$ occurs at time $t=t_0$ if and only if $y(t_0)=0$ and $\dot{y}(t_0)>0$, the analogous conditions holding in order to have the bounce in $B$. Hence, in general there can be a bounce in just one of the two scale factors.

According to the above-mentioned conditions, from Eqs. (\ref{eq.00})-(\ref{eq.22}) a bounce in $A$ requires
\begin{equation}
\Lambda > \frac{k \lambda}{2B(t_0)^4},
\end{equation}
i.e. the Strong Energy Condition of the total matter-energy content has to be violated. A similar analysis of  Eqs. (\ref{eq.00})-(\ref{eq.22}) shows that a bounce in $B$ is impossible. Indeed, Eqs. (\ref{eq.00})-(\ref{eq.22}) imply $2\dot{x}=-k\left(p_r+\rho\right)$; since the radial pressure and the energy density for the Skyrme fluid are related by $p_r=-\rho$ [see Eq.(\ref{pr})] this means that $x=0\Longleftrightarrow\dot{x}=0$.

Besides the bounce behavior, the condition for expanding or recollapsing solutions can be easily singled out. For istance, one can notice that subtracting Eq.(\ref{eq.11}) from Eq.(\ref{eq.00}) and assuming  $\dot{B}\neq 0$ (i.e. $x\neq0$), the relation between the two scale factor is trivial, in a sense:
\begin{equation} 
 \dot{B}=\text{const.} A\label{relabdot}.
\end{equation} 
Then, Eq.(\ref{eq.11}) only contains the function $B$ and the system reduces to two equations that read:
\begin{eqnarray}
\dot{x}&=&\frac{1}{2}\left[\frac{ k}{\tilde B}\left(1+\frac{\lambda}{2\tilde B}\right)+\Lambda-3x^2-\frac{1}{\tilde B}\right]\label{2dSys1}\\
\dot{\tilde B }&=&2 \tilde B x \label{2dSys2}.
\end{eqnarray} 
where
$ \tilde B=B^2$.
This system admits two fixed points in the $(x,\tilde B)$-plane, namely, $\mathcal{P}_{1}=(0,f_{-}(\Lambda,k,\lambda))$ and $\mathcal{P}_{2}=(0,f_{+}(\Lambda,k,\lambda))$ where
\begin{equation}
f_{\pm}(\Lambda,k,\lambda)=\frac{1 - k \pm \sqrt{1 - 2 k + k^2 - 2 k \lambda \Lambda}}{2\Lambda}
\end{equation}
For each point, there is always a pair of eigenvalues with opposite signs $(\lambda_{i},-\lambda_{i})$, namely
\begin{equation}
\lambda_{1,2}=\sqrt{\frac{-1 + k (2-k +2 \lambda \Lambda + S) \pm S }{k \Lambda}}
\end{equation}
with $S= \sqrt{(-1 + k)^2 - 2 k \lambda \Lambda}$.
By definition $\tilde B$ must be positive thus, one finds that the two fixed points exist in the following range:
\begin{equation}
0 < \lambda < \frac{1 - 2 k + k^2}{2 k \Lambda} \label{ParRange},
\end{equation}
$\mathcal{P}_{1}$ being neutrally stable, $\mathcal{P}_{2}$ being unstable of saddle type. 
We stress that for both the fixed points $\tilde B$ is constant, i.e. a constant scale factor $B$, thus the condition $\dot{B} \neq 0$ is violated. Moreover, Eq.(\ref{relabdot}) implies that, at the fixed points, the scale factor $A$ vanishes thus the cosmological model meets a singularity.

Interestingly enough, this analysis reveals a connection between the cosmological constant and the parameters of the Skyrme model. Considering the Skyrme parameters reported above and recalling that the cosmological constant value is $\Lambda\sim 10^{-52} m^{-2}$, one can deduce that the conditions on the model parameters in Eq.(\ref{ParRange}) are fulfilled for such estimations, thus for a large portion of the phase space, the solutions will evolve toward expansion.

\begin{figure}[th]
\begin{flushleft}
\includegraphics[scale=0.75]{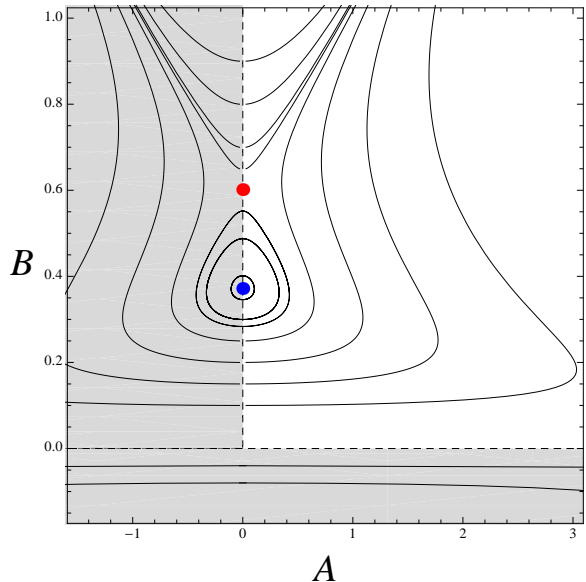}
\caption{Behavior of the scale factors in the $(A,B)$-plane. The parameters' values are arbitrarily chosen in the range Eqs.(13) as: $\Lambda =1, k = 0.5, \lambda = 0.2$. The upper (red) point and the lower (blue) point correspond to the fixed points $\mathcal{P}_{1}$ and $\mathcal{P}_{2}$ respectively.}
\label{ABp}
\end{flushleft}
\end{figure}

Thus, we distinguish two behaviors, namely, solutions in the basin of the stable fixed point $\mathcal{P}_{1}$ and unbounded solutions, which respectively correspond to the orbits in the basin of attraction of point $\mathcal{A}$ and $\mathcal{G}$ of the previous analysis. Their physical interpretation can be immediately understood considering the corresponding evolution of the scale factors $A$ and $B$  which can be readily determined by numerically integrating the system of Eqs.(\ref{eq.00}-\ref{eq.22}); the result presented in Fig.\ref{ABp}.
The system admits two type of solutions, recollapsing solutions belonging to a finite region containing the lower (blue) static solution (corresponding to the neutrally stable fixed point $\mathcal{P}_{1}$) and expanding solutions (influenced by the unstable fixed point $\mathcal{P}_{2}$). It is worth stressing that, being $A$ and $B$ two scale factors, the physical solutions are those living in the $A>0$ and $B>0$ region. Thus the actual behavior is the following: for initial conditions in the bounded region, there are solutions with expanding scale factor $B$ and recollapsing scale factor $A$. The other solutions, for generic initial conditions outside the bounded region, are characterized by an exponential expansion of both the scale factors at late time (see below). A very different behavior is observed in the case of a vanishing cosmological constant where only recollapsing solutions are present.

\subsection{Isotropization}

\begin{figure}[h]
\begin{flushleft}
\includegraphics[scale=0.65]{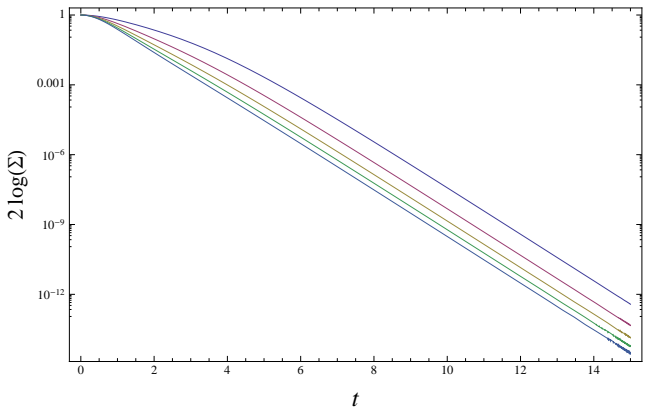}
\caption{Time evolution of the dimensionless shear parameter $\Sigma ^2$ for initial conditions outside the basin of the periodic solutions. The parameters' values are chosen to be the same as in Fig.\ref{ABp}. }
\label{isotropization}
\end{flushleft}
\end{figure}

\begin{figure}[htb]
\centering
\includegraphics[scale=0.75]{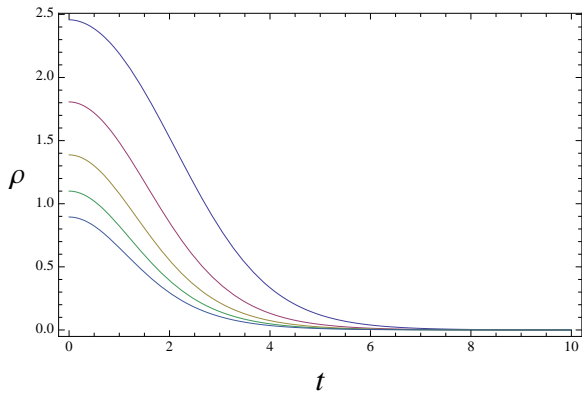}\\
\includegraphics[scale=0.75]{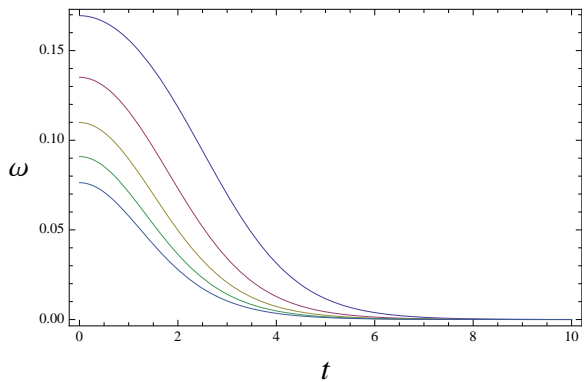}
\caption{{\protect\small {Time evolution of (a) the energy density and (b) the pressure parameters $\omega_{\theta}= \omega_{\phi}$ of the Skyrme fluid for initial conditions outside the basin of the recollapsing solutions. The parameters' values are chosen to be the same as in Fig.\ref{ABp}.}} }
\label{subfigs}
\end{figure}

For generic initial conditions outside the bounded, finite, stability basin dominated by the center fixed point $\mathcal{P}_{1}$ in Fig.\ref{ABp}, namely, for all the orbits converging to the attractor fixed point $\mathcal{G}$ of the previous analysis corresponding to de Sitter-like expanding solutions, the dimensionless shear parameter $\Sigma ^2$ undergoes an exponential decay. This means that the cosmological solutions exhibit isotropization within a finite amount of time, as one can easily see from Fig.\ref{isotropization}. It is worth stressing that, in this context, isotropization means a vanishing shear parameter, i.e. the two scale factors are characterized by the same functional dependence on time. Indeed, Kantowski-Sachs metrics are topologically inequivalent to Robertson-Walker models, the former having a four-dimensional isometry group, but no three-dimensional simply transitive subgroup, acting on the three space.  
The same results can be achieved using the equivalent definition of anisotropic parameter of the expansion given in \cite{Adhav:2011zzj}. A similar behavior is observed for the evolution of the energy density $\rho$ and anisotropic pressure parameters $\omega_i$ of the Skyrme fluid, derived in Eqs.(\ref{rho})-(\ref{pt}), see Fig.\ref{subfigs}.

The other isotropic fixed point is $\mathcal{B}$, that is a repeller, while other fixed points are anisotropic. Three of them are repellers ($\mathcal{H}$) or saddle ($\mathcal{C},\mathcal{F}$). This translates into the possibility to have both isotropic and anisotropic initial conditions or intermediate anisotropic conditions leading to structure formation, provided a sufficiently small anisotropy.

%%%%%%%%%%%%%%%%%%%%%%%%%%%%%%%%%%%%%%%%%%%%%%%%%%%%%%%%%%%%%%%%%%%%%%%%%%%%%%%%%%%%%%%%%%%%%%%%%%%%%%%%%%%%%%%%%%%%%%%%%%%%%%%%%%%%%%%%%%%%%%%%%%%%%%%%%%%%%%%%%%%%%%%%%%%%%%%%%%%%%%%%%%%%%%%%%%%%%%%%%%%%%%%
\section{Conclusions} \label{Conclusions}
We have considered the Kantowski-Sachs the cosmological model sourced by a Skyrme field and a cosmological constant in the framework of General Relativity. The hedgehog ansatz, together with the assumption of a constant radial profile function $\alpha=\pi/2$, allows to recast the stress-energy tensor of the Skyrme field as an anisotropic fluid whose contribution to the evolution of the solutions has been analyzed. 

We have shown that the cosmological equations can be reduced to a simple three-dimensional autonomous dynamical system with compact phase space. Three invariant two-dimensional submanifolds are found, namely $Q=\pm1$ and $\Omega_{\Lambda}=0$. In the region of the phase space corresponding to physical cosmological solutions, six isolated fixed points are found: two points being attractors, two points being repellers, the remaining four points being unstable of the saddle type. The functional dependence on time of the two scale factors $A$ and $B$ corresponding to each of these solutions has been reconstructed. The system also displays a normally hyperbolic equilibrium set.

A simple analysis shows that, while a bounce in the scale factor $B$ is impossible, a bounce in the scale factor $A$ is possible when the strong energy condition is violated.

Two types of late time behaviors are found, either anisotropic collapsing solutions of Kasner-like type, or exponentially expanding solutions  of de Sitter-like type.

%%%%%%%%%%%%%%%%%%%%%%%%%%%%%%%%%%%%%%%%%%%%%%%%%%%%%%%%%%%%%%%%%%%%%%%%%%%%%%%%%%%%%%%%%%%%%%%%%%%%%%%%%%%%%%%%%%%%%%%%%%%%%%%%%%%%%%%%%%%%%%%%%%%%%%%%%%%%%%%%%%%%%%%%%%%%%%%%%%%%%%%%%%%%%%%%%%%%%%%%%%%%%%%
\acknowledgements
The authors wish to thank Fabrizio Canfora for bringing the subject of the paper to their attention. This work is partially supported by Agenzia Spaziale Italiana (ASI) through Contract No. I/034/12/0. The authors acknowledge support by Istituto Nazionale di Fisica Nucleare (INFN) and by the Italian Ministero dell'Istruzione, dell'Universit\`{a} e della Ricerca (MIUR).

%%%%%%%%%%%%%%%%%%%%%%%%%%%%%%%%%%%%%%%%%%%%%%%%%%%%%%%%%%%%%%%%%%%%%%%%%%%%%%%%%%%%%%%%%%%%%%%%%%%%%%%%%%%%%%%%%%%%%%%%%%%%%%%%%%%%%%%%%%%%%%%%%%%%%%%%%%%%%%%%%%%%%%%%%%%%%%%%%%%%%%%%%%%%%%%%%%%%%%%%%%%%%%%

\end{document}